\documentclass[sort&compress,preprint,12pt]{elsarticle}
\journal{ }
\usepackage{amsmath,amssymb,graphicx,color,lipsum,mathrsfs}

\begin{document}

\begin{frontmatter}
\title{Competition May Increase Social Happiness in Bipartite Matching Problem}
\author[hyit,unifr]{Yi-Xiu Kong}
\author[shufe]{Guang-Hui Yuan}
\author[hyit]{Lei Zhou}
\author[unifr]{Rui-Jie Wu}
\author[hyit,unifr]{Gui-Yuan Shi\corref{cor1}}\ead{guiyuan.shi@unifr.ch}
\cortext[cor1]{Corresponding author}

\address[hyit]{College of Computer Engineering, Huaiyin Institute of Technology, Huaian 233003, PR China
}
\address[unifr]{Department of Physics, University of Fribourg, Fribourg 1700, Switzerland}
\address[shufe]{Fintech Research Institute, Shanghai University of Finance and Economics, Shanghai 200433,  PR China}

\begin{abstract}
Bipartite matching problem is to study two disjoint groups of agents who need to be matched pairwise. It can be applied to many real-world scenarios and explain many social phenomena. In this article, we study the effect of competition on bipartite matching problem by introducing correlated wish list. The results show that proper competition can improve the overall happiness of society and also reduce the instability of the matching result of unequal sized bipartite matching.
\end{abstract}

\end{frontmatter}

\section{Introduction}

Bipartite matching problem is to study that how the two disjoint groups of agents can be matched pairwise for their personal preferences, such as the matching between men and women, students and colleges, workers and jobs, consumers and products\cite{Gale1962, Roth1984, Roth1992} and many other scenarios\cite{Maggs2015,Lebedev2007,Hasan2015,Hitsch2010}. For convenience we use the paradigm of marriage problem, where N men and N women need be matched. In this problem, each participant is selfish, everyone tries to optimize their own choices and the competition is inevitable. A key question is how to find a stable solution in which there is no such a pair of man and woman who both prefer each other than the assigned partner\cite{Roth1982}. The number of stable solutions is very large\cite{Pittel1989, Dzierzawa2000}, the most famous one among them is obtained by the Gale-Shapley algorithm, which was awarded the Nobel Prize Economics in 2012.

In 1962, Gale and Shapley proved that for the same number of men and women, a stable solution can always be obtained through Gale-Shapley algorithm\cite{Gale1962}. In this algorithm, every agent has a wish list, which is a ranking list of all members from the opposite sex. Men act as suitors and send proposals to women according to their wish lists. When a woman has multiple candidate partners, she always retains the one she desires most. The algorithm will continue until all people find their spouse. This matching result can easily be proved stable, because the only way for men to improve their current situation is to send proposals to women who have rejected themselves, and the spouse of these women must be in front of this man in her list. Besides, it has also proved to be the men-optimal solution among all stable solutions.

Statistical physicists also find areas of interest in the bipartite matching problem because the model is very similar to a system in which two different particles interact\cite{Zhang2017, Chakraborti2015}.  We assume that everyone's satisfaction is corresponding to the ranking of her/his spouses in her/his wish list\cite{Omero1997}. It can be seen as a cost function, or energy in physicist terminology. If a person just happens to match with the person at the top of her/his wish list as a spouse, then he will be the happiest and have an energy of 1. In the worst case, one had to choose the person at the bottom of the list and the final energy was N. In most of the previous researches\cite{Dzierzawa2000, Omero1997, Shi2016}, for simplicity, the wish lists are always established randomly and independently. 

Physicists are always eager to find a solution with the lowest energy, which is often called the ground state\cite{Shi2016,Mezard1985,Mezard1987, Dotsenko2000}. The replica method in spin glass is used to compute the global optimal solution in the bipartite matching problem\cite{Mezard1985}, and the result gives that the average energy of each individual in the ground state is $0.808 \sqrt{N}$\cite{Omero1997}. However, 24.2\% of men can find one or more women\cite{Shi2016}, so that both of them prefer each other to their current spouse. Therefore, this ground state solution is very unstable.
Moreover, the Mean Field Theory is utilized to calculate the average energy of the stable solution obtained by Gale-Shapley algorithm\cite{Omero1997}. In this case, the average energy of men is $log(N)$, and the average energy of women is $N/log(N)$.  Further, with the aid of the energy distribution function, it is proved that for all stable solutions, there exists a relationship $\epsilon_m*\epsilon_w=N$. The average energy is $\epsilon_g =(\epsilon_m+\epsilon_w)/2 = 0.5×(\epsilon_m+N/\epsilon_m)$. It is easy to see that the global optimal stable solution corresponds to the same energy $\sqrt{N}$ for both men and women, which is a little worse than the global optimal solution. In addition, it is worthwhile to mention that the men-optimal solution has the highest total social energy among all stable solutions. After that, various issues were investigated, such as the matching problem under partial information\cite{Zhang2001,Laureti2003}, which means the lists of men only contain a fraction of all women, or people have a spatial distribution and tend to match with people who are geometrically closer, etc. 

One of the interesting questions is, if people's wish lists are correlated, which will induce competition among the agents on the same side, then how will the competition affect the matching results? In previous studies\cite{Caldarelli2001}, numerical simulation showed that the exerting of competition will lead to higher energy and lower happiness. Here we thoroughly study the impact of competition on bipartite matching, and instead of intuitive result that competition may reduce the well-being of the society, our result shows that proper competition can make society happier. On the other hand, a recent research\cite{Shi2018} shows that random bipartite matching is very unstable because reducing only one woman will lead to a dramatic change in the matching result. However, unstable situations are rarely observed in real life, even though the numbers of matching parties in reality are often different. Our research shows introducing competition can effectively increase the stability of the matching result and explain the absence of the unstable situation in daily observation.

\section{Method}
The Gale-Shapley algorithm\cite{Gale1962} is described below: suppose we have two disjoint sets of N men and N women who need to be matched pairwise. Each person has her/his wish list in which stores the ranking of all members from the opposite sex. At the beginning, everyone is unengaged. In each step, each unengaged man issues a proposal to his favorite woman among those he has not proposed to. The proposed woman will choose the her favorite from all the suitors and her provisional partner. The process iteratively runs until everyone is matched and it's easy to realize that this final matching will be achieved eventually.

\section{Result and Discusions}
\subsection{Matching between $N$ men and $N$ women}
Let us firstly consider the matching problem when the two groups have equal size. For simplicity, it is usually assumed that everyone's wish list is completely random. Considering the process of Gale-Shapley algorithm, men make proposals to women. If the proposed woman is unengaged, then the total number of matched pairs will increase by one. If this woman is engaged, no matter the suitor or her current partner is retained, the number of partners will not change.

It is proved that on average\cite{Omero1997}, a total number of $N(log(N)+0.522)$ proposals need to be sent, to make everyone engaged. Every proposal leads to a man's energy increasing by one, the average energy of men is the same as the average number of proposals that he needs to make, i.e. the average energy of man is $log(N)+0.522$. On the other hand, for a woman, each of them receives on average $log(N) + 0.522$ proposals. Each time they receive a proposal, they can make a choice and decide the man whom they prefer. Obviously, the greater the number of proposals, the happier they are. Since each suitor is randomly distributed in the wish list, the optimal one among many choices is equivalent to the first order statistic of multiple random sampling from even distribution. It is easy to obtain the woman's average energy is $N/(log( N) +0.522+1)$. When $N$ is large, the average male energy is approximately equal to $log(N)$, and the average female energy is approximately equal to $N/log(N)$. This conclusion is consistent with the results of the energy distribution function method used in previous study\cite{Omero1997}.

However, in reality the wish lists are often not random, some intrinsic properties, here expressed in fitness, will affect the ranking order of the wish lists. Beauty, intelligence, wealth, etc. can be widely viewed as high-fitness, people with those widely accepted attributes are easier to rank in front of the wish lists. In the extreme situation, all people's preferences are strictly based on fitness term, then everyone should have an identical wish list. In the beginning, all men make proposals to the woman with the highest fitness. The man with the highest fitness will be accepted and the others are refused. After that, the remaining men make proposals to the woman with the second highest fitness, the man with the second highest fitness will be accepted and the others were rejected. We continue this process, it is easy to know the men's energy from low to high are $1,2,3. . . N$ and the average energy (N+1)/2. The energy of women is the same.

For the general situation, that fitness has an impact on ranking, and the personal preferences are diversified so causes random deviations. We assume that woman $i$ rates man $\alpha$ with a score $S_{\alpha,i}$：
$$S_{\alpha,i} = \omega \times F_\alpha + (1-\omega) \times N_{\alpha,i}$$
Here $F_\alpha$ is fitness of man $\alpha$ and $N_{\alpha,i}$ is a random term. For simplicity, we assume that $F$ and $N$ are uniformly distributed on [0,1], and that weight of fitness $\omega$ is universal for all men and women.

The wish list of woman $i$ is generated according to the order of her ratings of all men. Similarly, we can obtain the wish list of everyone. With these wish lists, implementing Gale-Shapley algorithm, we will obtain the final stable matching.

\begin{figure}[htb]
\center\scalebox{0.35}[0.35]{\rotatebox{0}{\includegraphics{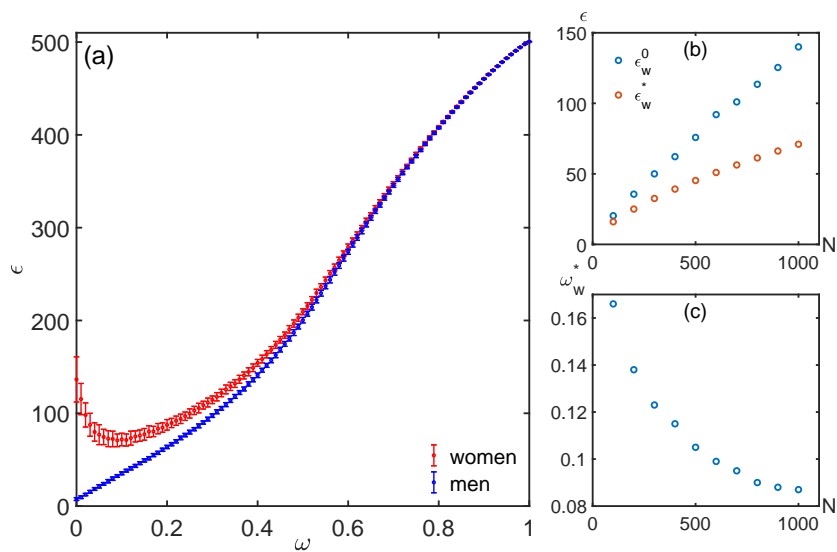}}}
\caption{
The numerical simulation of average energy of men and women, the result is averaged over 100 realizations. a) the average energy of men and women, as a function of $\omega$; b) the energy of w=0 and the energy of optimal $\omega$, denoted as $\omega_w^*$; c) the optimal weight $\omega_w^*$ versus $N$. 
}
\label{fig1}
\end{figure}

Now we start to analyze this matching result. As shown in Fig.1a, the average energy of men increases monotonically with the weight of fitness, $\omega$. This is because when the weight of fitness increases, it is easier for a woman with high fitness to be ranked in the front of the lists. When men make proposals according to their wish lists, their aims are often overlapping, which requires more proposals for everyone to be engaged. 

However, a slight competition will significantly reduce the women's energy. This improvement of women's happiness increases with $N$ (Fig.1b), and the competition required for the woman to reach optimal energy decreases as the population grows (Fig.1c). As competition further intensifies, the woman’s energy will also increase. On one hand, due to the increase of the number of proposals, the number of choices of woman increases. It can be known from the nature of the order statistics that the minimum number of multiple random sampling decreases as the number of sampling increases. The increase in competition between men is beneficial to the women. On the other hand, the increase in the weight of fitness also leads to women tend to favor the men with high fitness, and thus increases the intensity of competition among women. Many women have to be matched with the men positioned in the back in their wish lists. This will lead to an increase in the average energy of women and all people.

\begin{figure}[htb]
\center\scalebox{0.35}[0.35]{\rotatebox{0}{\includegraphics{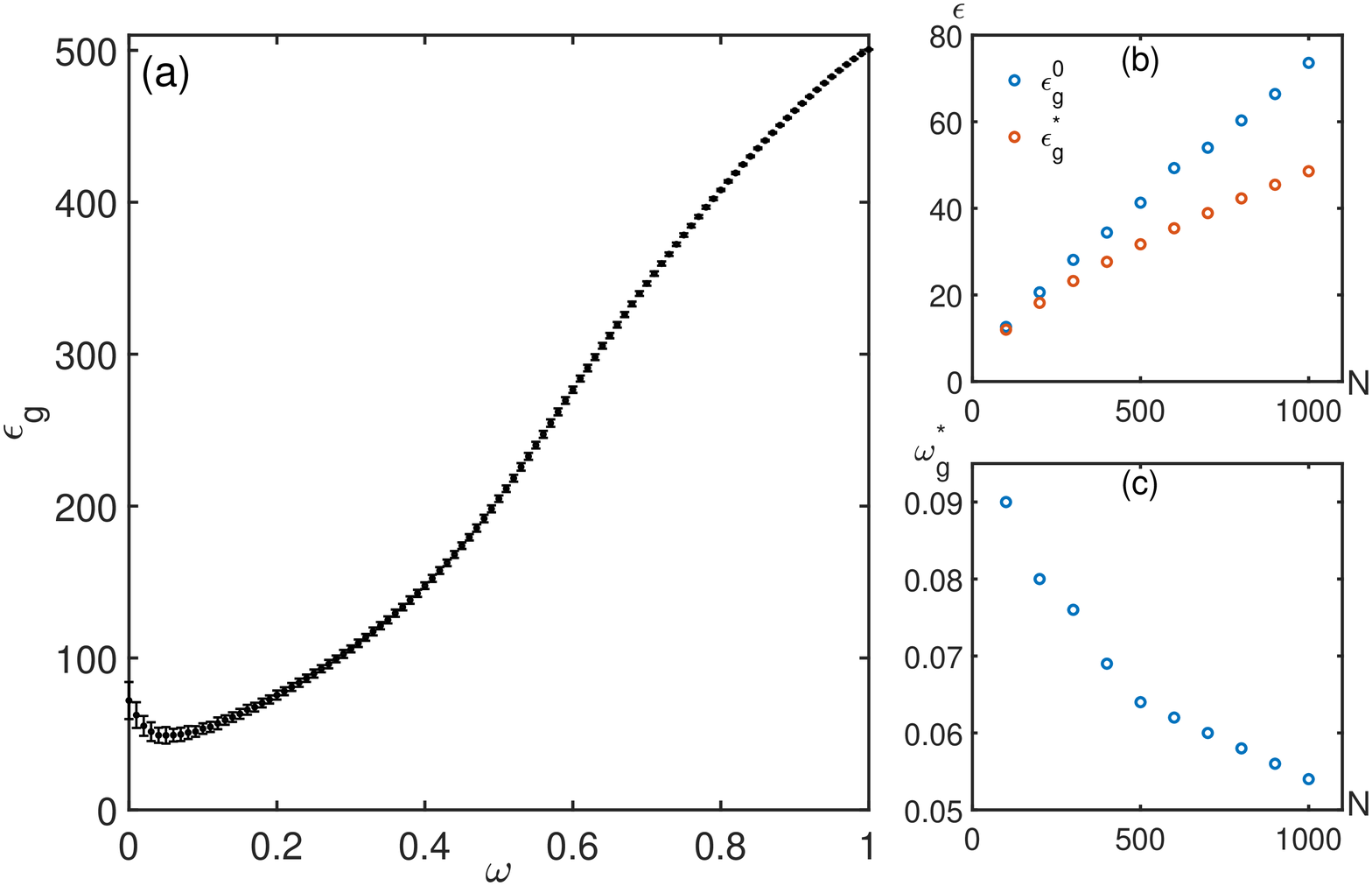}}}
\caption{
The numerical simulation of average energy of all people, the result is averaged over 100 realizations. a) the average energy of each person, with the change of $\omega$; b) the energy at $\omega = 0$ and the energy at optimal $\omega_g^*$; c) the optimal weight $\omega_g^*$ versus $N$.
}
\label{fig2}
\end{figure}

Due to the increase of $\omega$ in the beginning, the reduction of women's energy is more significant than the increase of men's energy. As shown in Fig. 2a, there exists a $\omega_g^*$ that can make average energy of all people reach the optimal value. In other words, a proper competition can help to increase the degree of social well-being.

In order to study the relationship between a person's intrinsic fitness and her/his level of happiness, we cluster the energy of men and women to 10 groups according to the range of fitness [0, 0.1), [0.1, 0.2), ..., [0.9, 1] (Data binning), and compare their happiness  in two cases: $\omega = 0$ and $\omega = \omega_w^*$ (women's optimal). As shown by Fig.3a, the competition causes energy of men in all fitness ranges to rise. As shown in Fig.3b, when competition is induced, the energy of women in lowest 20\% fitness groups increases, and the energy of remaining 80\% women decreases. In general, women will become happier with a certain level of competition, as we have known above. 

In addition, we take three representative agent groups and labeled as bottom (fitness ranges [0,0.1]), middle (fitness ranges [0.45,0.55]) and top (fitness ranges [0.9,1]) respectively. With the increase of competition, i.e. the increase of $\omega$, the energy of the three groups of men will increase, while the top women will be happier and the bottom women will be less happy (as shown in Fig.3c, d). This change is very sensitive to $\omega$, even if $\omega$ is very small (0.01), the happiness difference between the top group and bottom group has an obvious gap. For the middle group women, slight competition improves their situation, but as the competition intensifies, they become even more unhappy. 

\begin{figure}[htb]
\center\scalebox{0.5}[0.5]{\rotatebox{0}{\includegraphics{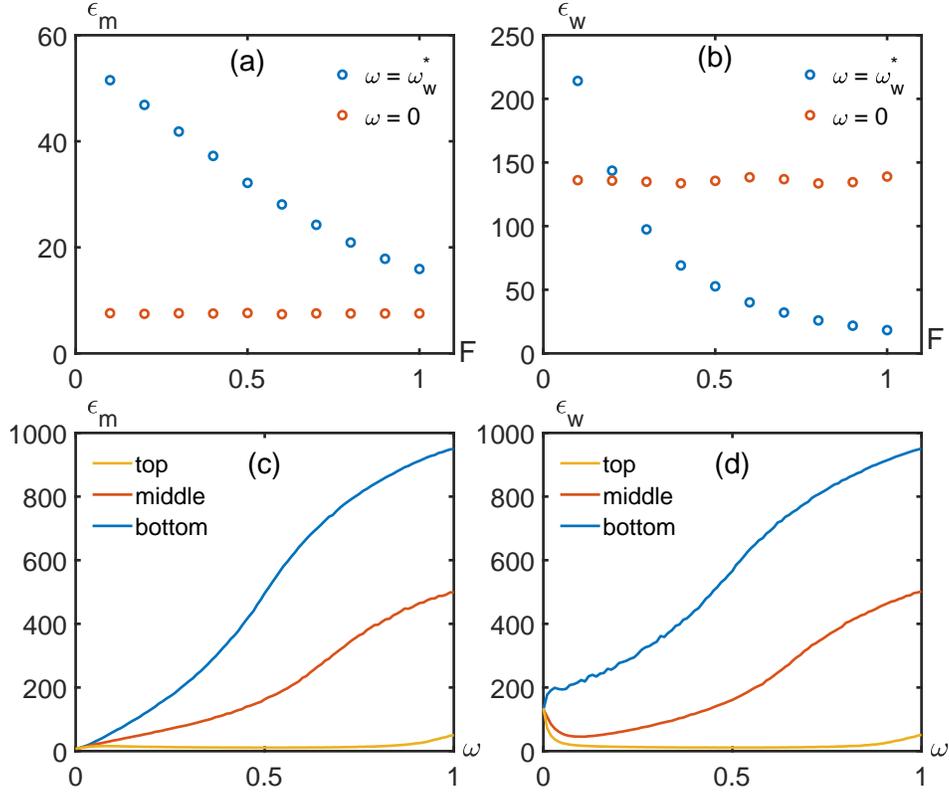}}}
\caption{
 The average energy of different fitness groups, the result is averaged over 100 realizations. a) the men's energy when $\omega = 0$ and $\omega = \omega_w^*$; b) the women's energy when $\omega = 0$ and $\omega=\omega_w^*$; c) the men's energy of three representative fitness groups, bottom(0,0.1), middle(0.45,0.55) and top(0.9,1), versus $\omega$; d) the women's energy of three representative fitness groups, bottom(0,0.1), middle(0.45,0.55) and top(0.9,1), versus $\omega$.
}
\label{fig3}
\end{figure}

\subsection{Matching between $N$ men and $N-1$ women}

For a long time, the G-S algorithm has been considered to produce a man-optimal stable matching. In particular, for a completely random wish list, that is when $\omega = 0$, the average energy of men is $log(N)+0.522$, which is far less than the average energy of women, which is $N/[log(N)+0.522]$, so the active side takes a huge advantage in the matching. However, a recent research\cite{Shi2018} shows that in a random bipartite matching, if one woman is removed from the matching, the average energy of men will become $N/log(N)$, and the average energy of women will become $log(N)$. The happiness of the positive and passive sides is completely reversed.

In reality, the numbers of matching parties are often not equal. Imagine reducing a member of passive party in two-side matching, such as reducing a woman in the marriage problem, or reducing a job position in the labor market, or reducing one offer in the college admission, under the assumption of random bipartite matching, the pairing result will change drastically with the smallest change of the size of passive side. But in reality, this situation is rarely seen , one of the possible reason is that, the low-fitness agents will be quickly eliminated, and other people's pairing results are almost unaffected.

We analyze the matching results in the case of 1000 men and 999 women with different $\omega$. As shown in Fig.4a, the curves are extremely similar to that of Fig.1a, despite the fact that the gender has reversed. The women become the donimant side of the matching. We compare this asymmetric case with the symmetric case above, and examine the changes of  the average energy of men and women , $\Delta\epsilon_m = \epsilon_m'-\epsilon_m$ and $\Delta\epsilon_w = \epsilon_w-\epsilon_w'$. The results are shown in Fig.4b and Fig.4c respectively. It is found that this difference sharply decreases with the increase of $\omega$, which explains why the slight change in the number of people in reality does not obviously affect the level of well-being in society.

\begin{figure}[htb]
\center\scalebox{0.35}[0.35]{\rotatebox{0}{\includegraphics{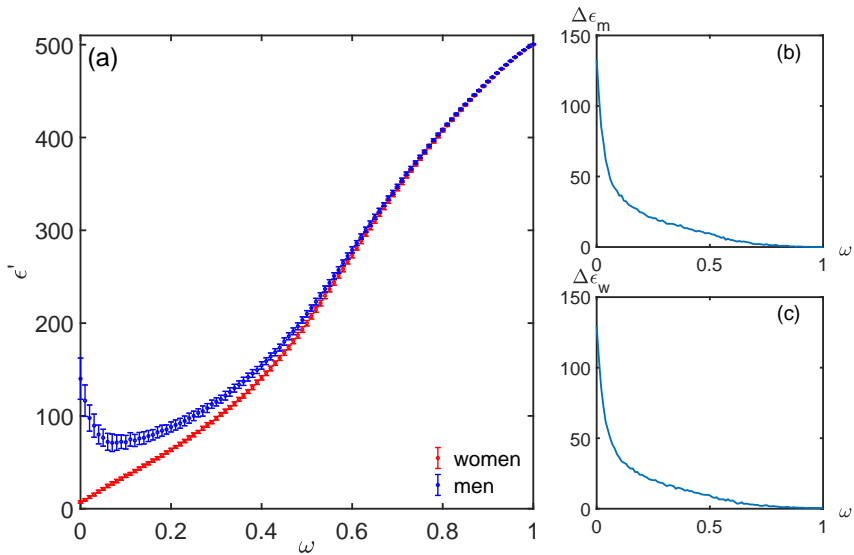}}}
\caption{
The matching simulation of 1000 men and 999 women, with different $\omega$, the result is averaged over 100 realizations. a) the average energy of men and women; b) $\Delta\epsilon_m$ versus $\omega$; c) $\Delta\epsilon_w$ versus $\omega$.
}
\label{fig4}
\end{figure}

\section{Conclusion}

In this article we introduce competition in bipartite matching problem by adopting correlated wish lists. While in the traditional model, everyone's wish list is not correlated, the active party can easily acquire the results they want and stop sending proposals so that the passive party are left with little freedom to choose. We show if a proper competition is introduced, men have to make more efforts and send more proposals, which will slightly reduce men's happiness. But at the same time, the happiness of women will be obviously increased, and society as a whole will become happier than the original matching result. In other words, proper competition increases the total happiness of the society. This is not only true for the bipartite matching problem, but also enlightens our understanding of many other social phenomena. Besides, the introduction of correlation in the wish lists can also significantly reduce the instability of matching results.

\section*{Conflict of Interest}

The authors declare no conflict of interest.

\section*{Acknowledgements}

We would like to thank Prof. Yi-Cheng Zhang and Wen-Yao Zhang for helpful discussions. This research is supported in part by the Chinese National Natural Science Foundation under grant No. 61602202 and Natural Science Foundation of Jiangsu Province under grants No. BK20160428, BK20161302. GYS and YXK acknowleges the support from China Scholarship Council (CSC).

\end{document}